# Bayesian Extreme Value Analysis of Stock Exchange Data

*Sean van der Merwe (Department of Mathematical Statistics and Actuarial Science, University of the Free State)*
*Darren Steven (Deloitte SA)*
*Martinette Pretorius*

**Abstract:**
The Solvency II Directive and Solvency Assessment and Management (the South African equivalent) give a Solvency Capital Requirement which is based on a 99.5% Value-at-Risk (VaR) calculation. This calculation involves aggregating individual risks. When considering log returns of financial instruments, especially with share prices, there are extreme losses that are observed from time to time that often do not fit whatever model is proposed for the regular trading behaviour. The problem of accurately modelling these extreme losses is addressed, which, in turn, assists with the calculation of tail probabilities such as the 99.5% VaR. The focus is on the fitting of the Generalized Pareto Distribution (GPD) beyond a threshold. We show how objective Bayes methods can improve parameter estimation and the calculation of risk measures. Lastly we consider the choice of threshold. All aspects are illustrated using share losses on the Johannesburg Stock Exchange (JSE).

## Introduction

The focus in this research is the quantification of risk. This is done through distributional modelling, particularly of log share returns. It is general knowledge that shares form a crucial part of most investment portfolios and investment strategies. The investor, knowing the volatile nature of shares, is often not overly concerned about minor fluctuations around the mean, but rather with the possibility of experiencing excessively large negative returns. This led to the exploration of the use of Extreme Value Theory (EVT) in financial modelling (see Klüppelberg, 2001 for an overview of this topic). This specific application of EVT has not been extensively researched, and the application of Bayesian methods in this context appears to be novel. To be clear, while the Bayesian approach to EVT is not new (see Smith, 2001 for two excellent examples) the application of this approach to log returns is. The most notable difference between the Bayesian and classical approaches can be seen in the calculation of risk measures, where parameter uncertainty is better taken into account.

Improvements in the calculation of risk measures are becoming more important with the increase in regulation with regard to capital requirements. In the South African insurance industry the Solvency Assessment and Management (SAM) directive includes a Solvency Capital Requirement which is based on a 99.5% Value-at-Risk (VaR) calculation. The directive is regularly updated by the Financial Services Board (FSB). See FSB (2014) for more information.

Many distributions have been used to model share returns as a whole – see Shao *et al.* (2001) for an overview – but we consider the method of Nascimento *et al.* (2011). They use the Generalized Pareto Distribution (GDP) above a threshold and combine it with a non-parametric approach below the threshold. The research of our paper, however, is not concerned with data below the threshold. We focus only on the extreme losses by taking the negative of the log returns and then choosing a positive threshold.

The choice of the GPD is not arbitrary. The theorem of Pickands (1975) states that, beyond an appropriate threshold, the extremes should fit the GPD, at least in an approximate sense.

Throughout this paper we will illustrate our results using either simulation studies or using the daily closing prices of the JSE Top 40 Index over a 10 year period (August 2002 to August 2012). For this data we select the threshold at a 1 day loss of 3.3%, as shown in Figure 1 below.

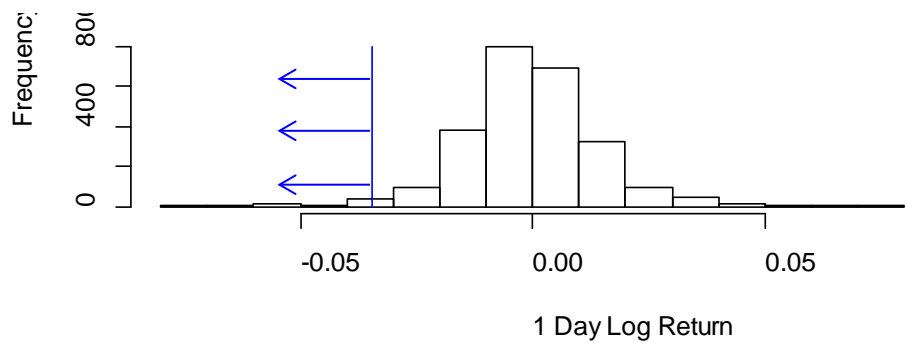

Figure 1: Log Returns of JSE Top 40 Index showing threshold.

The GPD was introduced by Pickands in 1975. The density function is given by:

$$f(x) = \frac{1}{\sigma}\left(1 + \frac{\gamma(x-\mu)}{\sigma}\right)^{-\frac{1}{\gamma}-1} \tag{1}$$

Note that $x \geq \mu$ when $\gamma \geq 0$ and $\mu < x < \mu - \frac{\sigma}{\gamma}$ when $\gamma < 0$.

Quantifying risk using this distribution requires three steps. The first is the choice of the threshold parameter $(\mu)$, which will be discussed in Section 2. The second is the estimation of the remaining parameters, which will be discussed in Section 3. The third is the calculation of risk measures, which will be discussed in Section 4. We then conclude by highlighting the advantages of this approach.

## 2. Threshold choice

Various methods have been proposed for choosing the threshold, the majority of which are visual and hence provide room for subjectivity. A sufficient threshold needs to be chosen in that it should be high enough to justify the use of the model (*i.e.* such that the GPD fits) but not too high as this would decrease the number of observations used and hence jeopardise the statistical reliability of the inferences.

One visual method suggested by Beirlant *et al.* (1996) makes use of a Pareto quantile plot and selects the threshold as the point at which the plot appears to become linear. Graphical tools such as the mean excess plot (see Beirlant *et al.,* 1996, Coles, 2001) and the Hill plot have also been used in the search for an optimal threshold.

Dupuis (1998) studied the choice of threshold with a method which uses robust estimation of parameters. An initial threshold is chosen and then parameter estimation is done. The estimation method assigns weights to the observations, with those observations which do not fit the applied model being assigned low weights. The threshold is then increased and the estimation repeated until all weights are close to one. This is one of the methods which aim to address the major issue of the visual methods, namely their subjectivity.

Another automated approach is presented by Thompson *et al.* (2009) which is based upon theory presented by Coles (2001). It involves the estimation of the parameters for each of a large number of possible thresholds. The process uses theory involving Normal distributions, and a threshold is chosen once Pearson's Chi-Squared test for Normality is satisfied.

Ultimately though, we recommend the method given in Verster *et al.* (2013) which combines the Bayesian estimation techniques, discussed in the next section, with the Kullback-Liebler divergence measure to choose a threshold such that an optimal fit is obtained in a Bayesian sense.

## 3. Parameter estimation

Once an appropriate threshold is chosen, we need to estimate the scale parameter ($\sigma$) as well as the extreme value index ($\gamma$). To simplify this process we first subtract the threshold from each observation, effectively making $\mu = 0$.

There is a vast amount of literature on the estimation of parameters of the GPD. Maximum Likelihood Estimation (MLE) appears to be the most widely used method (see Smith, 2001 for a description of this method). Two other popular approaches (described in Hosking and Wallis, 1987) are the probability weighted moments (PWM) method and the method of moments (MOM).

The extreme value index (EVI) is traditionally estimated by the well-known Hill and Pickands' estimators. When the number of upper order statistics used in these estimators is small, the variance of these estimators is large. When the number of upper order statistics used is great, however, an extensive bias is introduced. Peng (1998) provides an unbiased estimator of the EVI. This estimator, however, has a larger variance than the above estimators. These estimators are based on the Pareto distribution, not the GPD and we do not recommend their use.

From the literature it is clear that all of the available methods have various advantages and disadvantages. It is not always clear which method would be the best in different situations.

MLE is only valid for certain values of the parameters. Hosking and Wallis (1987) pointed out that the algorithms used in the calculation of the MLE may on occasion fail to converge. The MOM is a very simple approach but may result in increased sampling errors due to the squaring of observations. It is also known that MOM and PWM may provide estimates which are inconsistent with the observed data. This occurs when $\gamma < 0$ and sample values which are greater than the implied upper bound of $\frac{\sigma}{\gamma}$ are present (Castillo and Hadi, 1997).

Bermudez *et al.* (2001) puts forward the use of Bayesian methods in the peaks over threshold (POT) approach applied in this paper. To apply Bayesian methods a prior must be specified. The prior can take many forms. It is often characterized by an experienced expert. In the absence of expert knowledge, it is convenient to make use of so called objective priors, such as the Jeffreys prior (Jeffreys, 1961) and the maximal data information (MDI) prior (first mentioned in Zellner, 1971, pp. 41-53, and given explicitly in Zellner, 1997).

Beirlant *et al.* (2004) gives the MDI prior as

$$\pi(\sigma, \gamma) \propto \frac{e^{\gamma}}{\sigma} \tag{2}$$

This formula flows from the general case

$$\pi(\underline{\theta}) \propto \exp[E\{\log f(x|\underline{\theta})\}] \tag{3}$$

The log of the MDI prior is added to the log likelihood

$$\ell = \sum_{i=1}^{n} \ln f_X(x_i) = -n \ln \sigma - \left(\frac{1}{\gamma} + 1\right) \sum_{i=1}^{n} \ln\left(1 + \frac{\gamma x_i}{\sigma}\right) \tag{4}$$

To arrive at the log posterior

$$\ln p(\sigma, \gamma) = -(n+1) \ln \sigma - \left(\frac{1}{\gamma} + 1\right) \sum_{i=1}^{n} \ln\left(1 + \frac{\gamma x_i}{\sigma}\right) + \gamma + c \tag{5}$$

Note that $c$ is an unknown constant.

To estimate parameter values from this distribution we can obtain the posterior mode through numeric optimisation, or the posterior mean through simulation. The posterior mode is similar to the MLE and should be equal to the MLE if a uniform prior is used or the sample size is large.

The posterior mean attempts to make use of the whole posterior instead of just the maximum, but requires the simulation of a large sample from the posterior distribution since analytic integration of the posterior is not feasible. We implement the simulation using the Metropolis algorithm in its simplest form (Metropolis *et al.*, 1953):

Given an observation $\underline{\theta}_1$ from this distribution, accept $\underline{\theta}_2$ as being from this distribution if

$$\frac{p(\underline{\theta}_2)}{p(\underline{\theta}_1)} > U_i \sim Uniform(0,1) \ or \ \ln p(\underline{\theta}_2) - \ln p(\underline{\theta}_1) > \ln U_i \tag{6}$$

Note that there is no need to calculate the value of the constant ($c$) as it cancels in the algorithm. The candidate observations $\left(\underline{\theta}_2 = (\sigma_2, \gamma_2)\right)$ are chosen from a bivariate Normal distribution in this case, with parameter values picked for good convergence properties.

A simulation study is performed in order to compare some of the different estimators discussed in this paper. Sample sizes of $n = \{40, 80, 120\}$ are used, along with the following values of $\gamma: \{-0.2, 0.3, 0.8\}$. Results are based on 5000 simulation runs for each combination of $n$ and $\gamma$. An additional simulation with 5000 runs with $n = 120$, $\gamma = 0.3$, and $\sigma = 0.008$ is also done to demonstrate that the methods work with typically observed values of negative differenced log share returns from individuals shares.

If we consider the observed root mean square errors (RMSE) and average them over the runs then we obtain results as in Table 1. It is clear that the Bayesian estimators perform best in almost all cases.

| Combinations | | MOM | PWM | MODE | | MEAN | |
|---|---|---|---|---|---|---|---|
| | | | | MDI | JEFF | MDI | JEFF |
| $n = 40$ | | | | | | | |
| 1 | σ = 1 | 0.238 | 0.243 | 0.254 | 0.236 | 0.244 | 0.243 |
| | γ = - 0.2 | 0.183 | 0.191 | 0.190 | 0.187 | 0.185 | 0.181 |
| 2 | σ = 1 | 0.292 | 0.287 | 0.283 | 0.290 | 0.288 | 0.285 |
| | γ = 0.3 | 0.235 | 0.229 | 0.228 | 0.234 | 0.230 | 0.235 |
| 3 | σ = 1 | 7.962 | 39.477 | 6.328 | 5.639 | 21.068 | 4.396 |
| | γ = 0.8 | 0.340 | 0.346 | 0.340 | 0.347 | 0.349 | 0.338 |
| $n = 80$ | | | | | | | |
| 1 | σ = 1 | 0.167 | 0.164 | 0.165 | 0.159 | 0.165 | 0.163 |
| | γ = - 0.2 | 0.1225 | 0.1238 | 0.1253 | 0.1221 | 0.1217 | 0.1245 |
| 2 | σ = 1 | 0.1911 | 0.1912 | 0.2034 | 0.2021 | 0.2038 | 0.1906 |
| | γ = 0.3 | 0.154 | 0.160 | 0.163 | 0.159 | 0.161 | 0.156 |
| 3 | σ = 1 | 2.024 | 1.176 | 1.832 | 1.226 | 0.964 | 6.111 |
| | γ = 0.8 | 0.257 | 0.255 | 0.251 | 0.259 | 0.254 | 0.257 |
| $n = 120$ | | | | | | | |
| 1 | σ = 1 | 0.133 | 0.129 | 0.129 | 0.124 | 0.134 | 0.129 |
| | γ = - 0.2 | 0.0992 | 0.0947 | 0.0945 | 0.0942 | 0.0962 | 0.0941 |
| 2 | σ = 1 | 0.166 | 0.152 | 0.150 | 0.160 | 0.160 | 0.151 |
| | γ = 0.3 | 0.1268 | 0.1229 | 0.1253 | 0.1231 | 0.1250 | 0.1243 |
| 3 | σ = 1 | 3.091 | 1.827 | 1.487 | 1.437 | 1.081 | 1.636 |
| | γ = 0.8 | 0.2217 | 0.2237 | 0.2211 | 0.2280 | 0.2219 | 0.2197 |
| 4 | σ = 0.008 | 0.00125 | 0.00124 | 0.00128 | 0.00123 | 0.00122 | 0.00118 |
| | γ = 0.3 | 0.123 | 0.124 | 0.127 | 0.122 | 0.125 | 0.126 |

Table 1: RMSE values based on simulation study of 10x5000 samples. The highlighted values are the lowest in each row.

For the JSE data we take a sample of size 10,000 from the joint posterior of the parameters. Figure 2 below illustrates this sample by means of a scatter plot. We can obtain parameter estimates, if we wish, by averaging over this sample. More importantly, we can obtain intervals for the parameters and quantify the inherent uncertainty in their estimation.

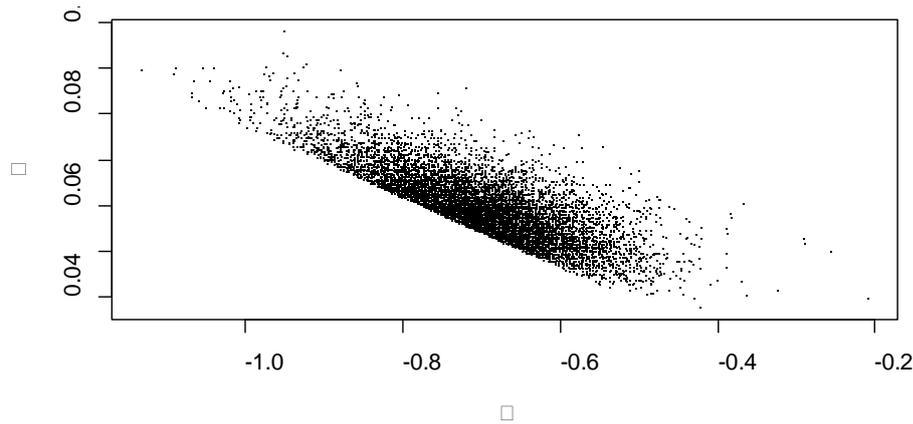

Figure 2: Scatterplot from joint posterior for extremely losses on the JSE Top 40 Index.

## 4. Risk measures

Value at risk (VaR) is a very popular measure for risk in a financial context. "The VaR summarizes the worst loss over a target horizon with a given level of confidence and summarizes the overall market risk faced by an institution." (Dowd 1998, Jorion 1997, cited in Assaf 2009).

The $(1-\alpha)100\%$ VaR is the amount of loss expected to occur $\alpha 100\%$ of the time during a given period. As an example, the 99% VaR can be interpreted as the loss expected to occur only once in 100 days.

In their article Hoogerheide and van Dijk (2008) illustrate a method for calculating VaR and Expected Shortfall (ES) in a Bayesian framework. They referred to the $100\alpha\%$ VaR as the $100(1-\alpha)\%$ quantile of the percentage return's distribution. They refer to the ES as the expected percentage return given that the loss exceeds the $100(1-\alpha)\%$ quantile.

The historical approach tries to estimate these measures based on the empirical distribution of the data. Unfortunately, since the data is always limited in some way, this approach cannot work for very high quantiles. The historical approach appears to stagnate and underestimate extreme risk.

The GPD on the other hand continues to infinity and only requires one to know the values of the parameters. The values of these parameters are not known with certainty in practice but the Bayesian framework allows us to quantify this uncertainty. Explicitly, we make use of the posterior predictive distribution:

$$p(x^f|\underline{x}) = \int_{\underline{\theta}} p(x^f|\underline{\theta})\, p(\underline{\theta}|\underline{x})d\underline{\theta} \qquad (7)$$

If uncertainty about an unknown parameter is captured in a posterior distribution, a predictive distribution for any quantity $z$ that depends on the unknown parameter, through a sampling distribution, can be obtained by the above mentioned formula.

In this case $p(x^f|\theta)$ refers to a new GPD observation given a set of parameters. To simulate from the GPD we can use the following transformation:

$$U \sim Uniform(0,1) \Rightarrow X = \left[(U^{-\gamma} - 1)\frac{\sigma}{\gamma} + \mu\right] \sim GPD(\mu, \sigma, \gamma) \quad (8)$$

We could thus simulate a large number of large samples and calculate the risk measures for each. Luckily, in the case of the GPD, both the VaR and ES have closed form expressions and simulation is not necessary. Explicitly:

$$VaR(1 - \alpha) = (\alpha^{-\gamma} - 1)\frac{\sigma}{\gamma} + \mu \quad (9)$$

$$ES(1 - \alpha) = VaR(1 - \alpha) + \frac{\sigma \alpha^{-\gamma}}{1 - \gamma} \quad (10)$$

These measures are ordered to obtain quantiles for the purpose of creating intervals. Note that since we are dealing with negative log share returns, which are GPD above a suitable threshold, it is necessary to rescale the alpha by multiplying by $\frac{\text{number of observations}}{\text{number of exceedences}}$.

Below is the graph for the VaR using the Bayesian methodology, rescaled to show the value that would be expected once in a given number of trading days. The historical method is imposed on that. Also given are the risk measures calculated on the assumption of Normality. The graphs clearly highlight the danger of this assumption.

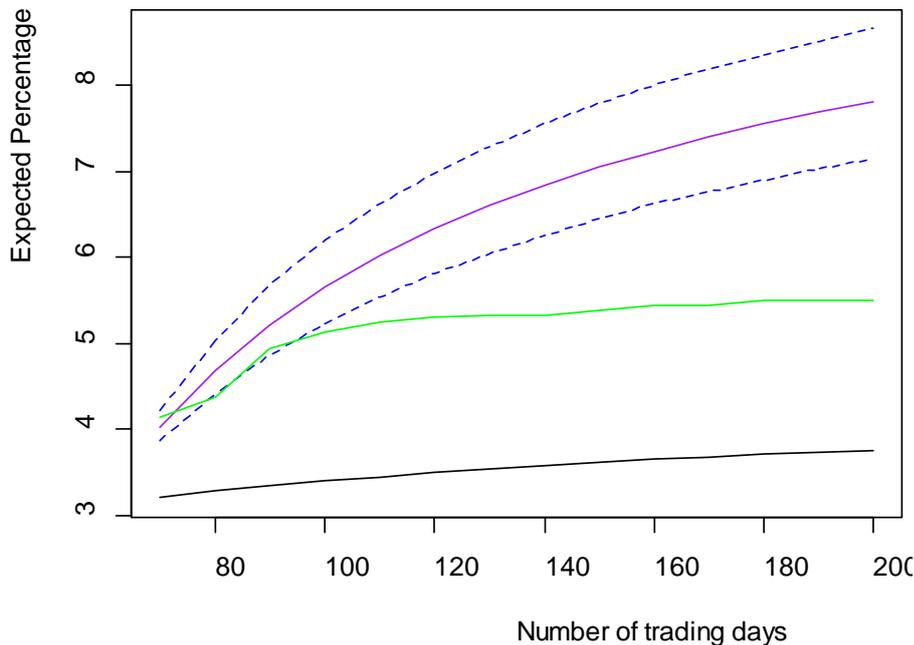

Figure 1: Bayesian VaR once per $n$ trading days (smooth purple) with 95% bounds (blue dotted) and historical approach (green stepped). Also given is the VaR assuming a Normal distribution (black curve at the bottom).

On the far right of Figure 1 is the 99.5% VaR. There is a clear difference between the approaches applied to this data. The same pattern applies to individual companies. Figure 2 shows the 99% VaR to 99.5% VaR for City Lodge Hotels Ltd. over the same period.

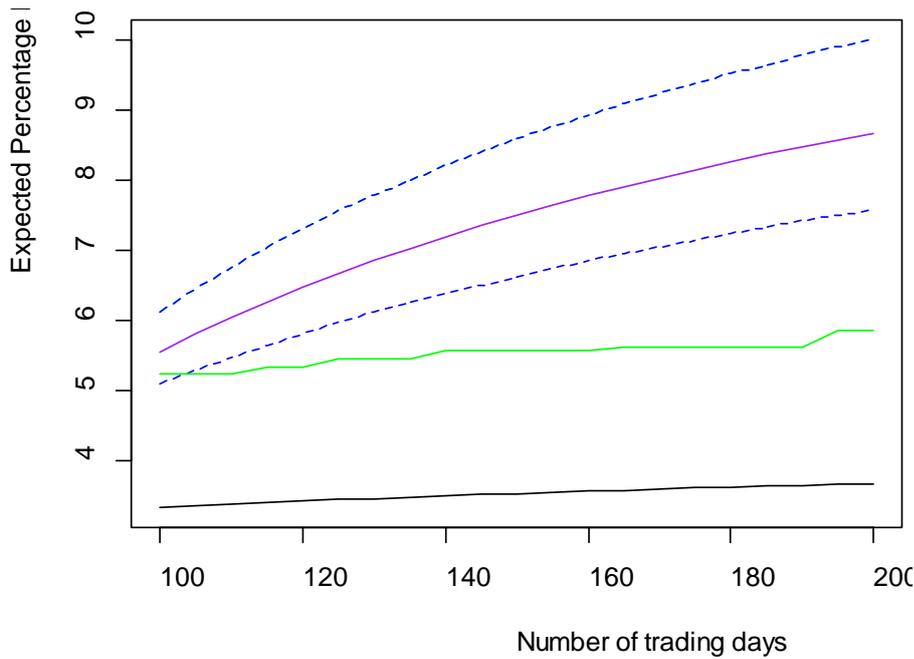

Figure 2: 99% VaR to 99.5% VaR for City Lodge Hotels Ltd. Bayesian VaR once per $n$ trading days (smooth purple) with 95% bounds (blue dotted), historical approach (green stepped) and VaR assuming a Normal distribution (black curve at the bottom).

Let us now consider the uncertainty in the choice of threshold. It is straightforward to repeat the procedures above for a selection of thresholds. In this case the original choice included very few observations, so if we include additional observations in the estimation procedure then the lines in the graph shift slightly. We would expect that using more observations will decrease the variance of the estimates, bringing the upper and lower bounds closer together; but at the same time it may increase the bias (increasing the chance of underestimating or overestimating the extreme losses).

This is indeed what we observe if we consider Figure 3 below. The saturated lines are the original estimates and the lines fade out as more observations are included.

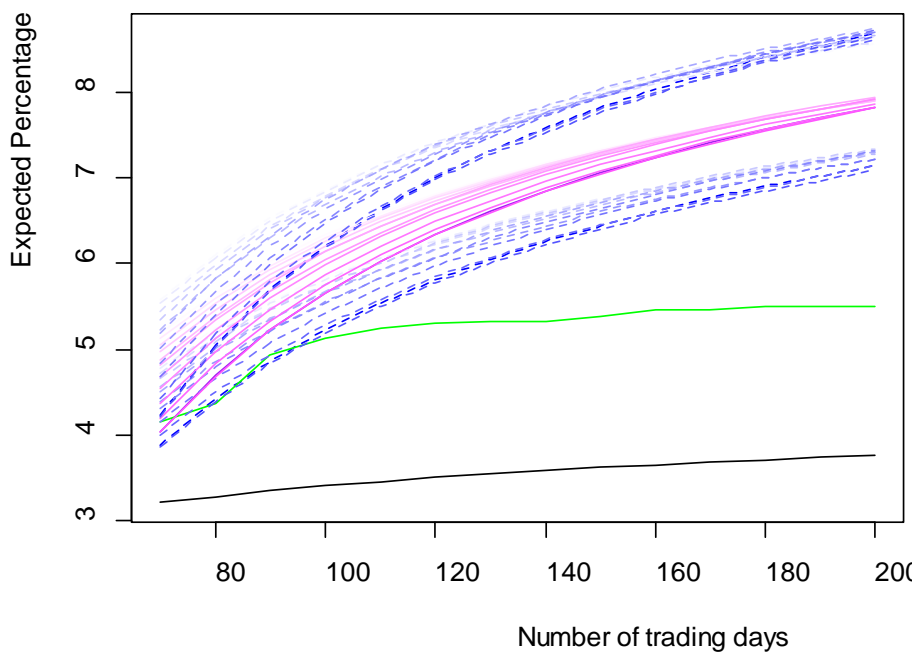

Figure 3: Bayesian VaR once per $n$ trading days (smooth purple) with 95% bounds (blue dotted), historical approach (green stepped) and Normal model (black curve at the bottom). Increased saturation corresponds with more extreme threshold.

## 5. Conclusion

This paper highlighted three ways in which the Bayesian framework assists us in modelling extreme losses that can occur in share data: First, the identification of an appropriate threshold beyond which it appropriate to apply extreme value theory. Second, the estimation of parameters and especially the ability to quantify the uncertainty in the parameter estimates. Lastly, the ability to estimate risk measures far into the tail with a clear indication of the accuracy of these measures.

Further, this paper highlighted the large discrepancy between the expected losses based on the Normal model, the expected losses obtained by extrapolating the observed frequencies, and the losses expected by extreme value theory. This can have serious implications with respect to capital requirements.

*Note: All techniques applied in this paper were implemented in the free open-source statistical package R (R Core Team, 2013). The classical extreme value analysis is done using the package evir (Pfaff and McNeil, 2012). Bayesian methods are coded in custom functions, available on request from the corresponding author.*